\documentclass[preprint,showpacs,amssymb,aps,pra]{revtex4}

\def\diagram#1{{\normallineskip=8pt
       \normalbaselineskip=0pt \matrix{#1}}}

\def\diagramrightarrow#1#2{\smash{\mathop{\hbox to 
.8in{\rightarrowfill}}
        \limits^{\scriptstyle #1}_{\scriptstyle #2}}}

\def\diagramleftarrow#1#2{\smash{\mathop{\hbox to .8in{\leftarrowfill}}
        \limits^{\scriptstyle #1}_{\scriptstyle #2}}}

\def\diagramdownarrow#1#2{\llap{$\scriptstyle #1$}\left\downarrow
    \vcenter to .6in{}\right.\rlap{$\scriptstyle #2$}}

\begin{document}

\title{Renormalized path integral for the 
\\
two-dimensional ${\boldmath \delta}$-function interaction}
                                                                               
\author{
Horacio E. Camblong$^{1}$
and
Carlos R. Ord\'{o}\~{n}ez$^{2,3}$}

\affiliation{$^1$ 
\mbox{Department of Physics, University of San Francisco, San Francisco, CA 94117-1080} 
\\
$^2$ Department of Physics, 
University of Houston, Houston, TX 77204-5506
\\
$^3$
World Laboratory Center for Pan-American Collaboration in Science and
Technology,
\\
University of Houston Center, Houston, TX 77204-5506}

\begin{abstract}
A path-integral approach for
$\delta$-function potentials is presented.
Particular attention is paid to the two-dimensional case,
which illustrates the realization of a quantum anomaly for
a scale-invariant problem in quantum mechanics.
Our treatment is based on an infinite summation of perturbation 
theory that captures the nonperturbative nature of the
$\delta$-function bound state. 
The well-known singular character of the 
two-dimensional $\delta$-function potential is dealt 
with by considering the renormalized path integral
resulting from a variety of schemes:
dimensional, 
momentum-cutoff,
 and real-space regularization.
Moreover, compatibility of the bound-state and scattering sectors is shown.
\end{abstract}

\pacs{03.65.Db, 11.10.Gh, 11.10.-z, 11.10.St}

\maketitle

\section{Introduction}

It is well known that the study of singular 
potentials in quantum mechanics leads to an array 
of technical difficulties~\cite{fra:71}.
In the past few years the relevance of field-theory tools for the 
analysis of singular potentials 
has been gradually recognized and successfully applied
to a number of interesting cases.
However, most of the published research to date
has been performed  in the Schr\"{o}dinger picture
by directly renormalizing the solutions obtained from
the Schr\"{o}dinger equation.
In contrast with these earlier works,  we have recently advanced a broad 
program that includes as a major objective
the treatment of these potentials within a path-integral framework~\cite{cam:pi_bs}.
The successful completion of this goal should
provide the first natural step in 
our program, the ultimate purpose of which
is to tackle the notoriously
difficult problem of bound states in quantum field 
theory~\cite{bs_in_qft,sav:99}.
The relevance of 
singular potentials is highlighted by their 
inevitable presence in the nonrelativistic reduction from field
theory to quantum mechanics~\cite{Stevenson,Beane} within the effective-field-theory 
paradigm~\cite{recent_review}.
In particular, in Ref.~\cite{cam:pi_bs}
we were able to show
explicitly that: (i) {\em infinite summations of perturbation theory \/}
capture the intrinsic nonperturbative nature of bound 
states~\cite{bha:89,gro:98};
(ii) the use of this technique and 
a {\em renormalization analysis\/}~\cite{gup:93,cam:ISP_prl,cam:dtI,cam:dtII}
provide a path-integral rederivation of the solution to the
inverse square potential~\cite{mot:49}.
In this paper we now show that a similar approach
reproduces the behavior of 
the two-dimensional $\delta$-function 
potential,
another apparently pathological
case~\cite{tho:79,hua:92,jac:91}.
A central result of this research is the {\em predictive power\/} of the general framework:
once renormalization is implemented at the level of the bound state(s) of the theory,
the scattering observables are uniquely determined.

The two-dimensional $\delta$-function
interaction is actually included in a larger class
of singular potentials whose phenomenological usefulness
dates back to the introduction of 
pseudopotentials
in the  early days of quantum
mechanics~\cite{bet:35}, and  the subsequent applications of the
zero-range potential in nuclear physics~\cite{bre:47,kru:01}, condensed-matter 
physics~\cite{kos:54}, statistical mechanics~\cite{lie:63},
 atomic physics~\cite{dem:89}, and particle physics~\cite{tho:79}.
Our understanding of these singular problems has advanced
considerably in recent times~\cite{zel:60}:
with the modern theory of pseudopotentials~\cite{alb:88,wod:91},
which overlaps with the technique of self-adjoint
extensions~\cite{jac:91,gol:77,ger:89a};
with the study of the nonrelativistic
limit of the $\phi^{4}$ theory and the 
question of its triviality~\cite{beg:85}; and
with the application of quantum-field-theory tools
to singular problems in quantum mechanics.
In particular, there have been many renormalization
analyses of the two-dimensional  $\delta$-function potential:
using momentum-space 
regularization~\cite{tho:79,hua:92,jac:91,phi:98a,mit:98,hen:98,nye:00},
as well as 
a momentum-space renormalization-group analysis~\cite{adh:95a};
using real-space regularization~\cite{man:94,gos:91,mea:91,per:91};
using dimensional 
regularization~\cite{cam:dtI,cam:dtII,phi:98a,mit:98};
and dealing with its associated quantum 
anomaly~\cite{jac:91,hol:93,cab:96}.

In contrast with the research published to date, 
in this paper we deal with the renormalization
of the two-dimensional  $\delta$-function potential
by using path integrals and infinite summations of perturbation theory from the outset.
In Sec.~\ref{sec:framework} we introduce the general framework, followed by
a presentation of the $\delta$-function interaction in
Sec.~\ref{sec:Delta-Function_Interactions}.
Renormalization of the bound-state sector is analyzed 
within  dimensional regularization in Sec.~\ref{sec:DR},
momentum-cutoff regularization in Sec.~\ref{sec:momentum_cutoff},
 and real-space regularization in Sec.~\ref{sec:real_space_cutoff}.
This is  followed by an analysis of the scattering sector
in Sec.~\ref{sec:scattering} and conclusions in Sec.~\ref{sec:conclusions}.
Finally, the appendices summarize the derivation and properties of
the free-particle Green's functions used throughout the paper.

\section{General Framework: Infinite Summations of Perturbation Theory}
\label{sec:framework}

Our stated goal of renormalizing the $\delta$-function interaction is best accomplished by first 
establishing the general framework.
We will now focus on the general technique of infinite summations
of perturbation theory 
and start by introducing the path-integral framework in its different
formulations needed for subsequent derivations. 

\subsection{General definitions}

The central object in the path-integral treatment for a 
nonrelativistic particle of mass $M$ 
is the quantum-mechanical propagator
\begin{eqnarray}
K_{D}({\bf r}^{''}, {\bf r}^{'}; t^{''},t^{'}) 
& = &
\int_{  {\bf r} (t')  = {\bf r}'  }^{  {\bf r} (t'')  = {\bf r}'' }
 \;  
{\cal D} {\bf r} (t) \,
\exp \left\{ 
\frac{i}{\hbar} 
S \left[ {\bf r}(t)  \right]  ({\bf r}'', {\bf r}' ; t'',t')  
\right\}
\label{eq:propagator_QM}
\\
& = &  
\lim_{N \rightarrow \infty}
\left( \frac{M}{2 \pi i \epsilon \hbar} \right)^{DN/2}
\,
\left[
\prod_{k=1}^{N-1} \int_{ \mathbb{R}^{D} } d^{D} {\bf r}_{k} \right]
\,
\nonumber \\
& \times &
\exp \left\{
\frac{i}{\hbar} \sum_{j=0}^{N-1} 
\left[
\frac{ M
( {\bf r}_{j+1}
- {\bf r}_{j}  )^{2} }{2\epsilon} 
-
\epsilon V({\bf r}_{j}, t_{j})
\right]
\right\}
\;  ,
\label{eq:propagator_QM2}
\end{eqnarray}
where the interaction with
a potential $V({\bf r},t)$ is considered
in the generic $D$-dimensional case.
In the continuous version, Eq.~(\ref{eq:propagator_QM}),
$S \left[ {\bf r}(t)  \right]  ({\bf r}'', {\bf r}' ; t'',t')  $ 
stands for the classical action associated 
with paths $ {\bf r}(t)  $ connecting the end points
${\bf r} (t')  = {\bf r}'$ and $  {\bf r} (t'')  = {\bf r}'' $.
In the time-lattice version, Eq.~(\ref{eq:propagator_QM2}),
a time slicing 
\begin{equation}
t_{j}= t'+j \epsilon
\;  ,
\; \; \; \; \; \; \; \;
\epsilon=
\frac{(t^{''}-t^{'})}{N}
 \; \; \; \; \; \; \; \;
(j=0, \cdots,N)
\; 
\end{equation}
 is introduced, with the additional notation
$ {\bf r}_{j} = {\bf r} (t_{j})$, while 
 the end points satisfy
$t_{0} \equiv t'$, $t_{N} \equiv t''$, 
${\bf r}_{0} \equiv {\bf r'}$,
and 
${\bf r}_{N} \equiv {\bf r''}$.
In Ref.~\cite{cam:pi_bs}
we showed how to derive the general expressions for the required infinite summations 
by starting with the basic lattice definition~(\ref{eq:propagator_QM2});
these results are reviewed below.

For the particular case of a  time-independent potential---which applies to
the $\delta$-function interactions of this paper---Eq.~(\ref{eq:propagator_QM}) depends upon the 
times $t'$ and $t''$ of the end points
only through their difference $T=t^{''}-t^{'} $,
so that we write $ K_{D} ({\bf r''},{\bf r'};t'',t') \equiv K_{D} ({\bf r''},{\bf r'};T)$.
Then, a complete analysis of the spectrum
is most effectively carried out by introducing the energy
Green's function 
$G_{D} ({\bf r''},{\bf r'};E)$
as the Fourier transform with respect to $T$ of the retarded Green's function
$ G_{D} ({\bf r''},{\bf r'};T)
= 
K_{D} ({\bf r''},{\bf r'};T) \,
\theta (T)$,
where  $\theta (T)$ stands for the Heaviside function.
Explicitly~\cite{cam:pi_bs,kleinert},
\begin{eqnarray}
G_{D} ({\bf r''},{\bf r'};E) 
& = &
\frac{1}{i\hbar}
\int_{0}^{\infty} dT  e^{iET/\hbar}
K_{D}({\bf r}^{''}, {\bf r}^{'}; T)
\label{eq:GreenF_from_Fourier_transform}
\\
& = &
\frac{1}{ i\hbar }
\int_{0}^{\infty} dT  
\,
\int_{  {\bf r} (0)  = {\bf r}'  }^{  {\bf r} (T)  = {\bf r}'' }
 \;  
{\cal D} {\bf r} (t) \,
\exp 
\left(
\frac{i}{\hbar} 
\left\{ 
S \left[ {\bf r}(t)  \right]  ({\bf r}'', {\bf r}' ; T)  +
ET 
\right\}
\right)
\; 
\label{eq:GreenF_from_PI}
\end{eqnarray}
is defined
so that
\begin{equation}
G_{D} ({\bf r''},{\bf r'};E) 
=
\left\langle 
{\bf r}'' 
\left|
\left( 
E - \hat{H} + i \delta 
\right)^{-1}
 \right|
{\bf r}'
 \right\rangle
\; ,
\label{eq:GreenF_operator}
\end{equation}
which is the usual definition in terms of the Green's function operator,
with $\delta = 0^{+}$.

\subsection{Perturbation theory and infinite summations}

A thorough analysis of the problem at hand can be completed
by means of perturbative expansions, as we shall discuss next.
At first sight one might dismiss this approach because it
 is well known that
finite summations of perturbation theory are unable to
capture the required
nonperturbative behavior. 
However,
this limitation can be removed when
infinite summations are considered.
The technique of infinite
perturbative summations~\cite{bha:89} has already been successfully
applied to a number
of interesting problems~\cite{cam:pi_bs,gro:98,gro:93}. 
Let us now summarize 
the main results of this approach, as needed for the present paper.

The perturbation expansion
starts  with the resolution of the action
\begin{eqnarray}
S 
\left[ {\bf r}(t)  \right]  ({\bf r}'', {\bf r}' ; t'',t')  
& = &
S^{(0)}   
\left[ {\bf r}(t)  \right]  ({\bf r}'', {\bf r}' ; t'',t')  
+
\tilde{S}  
\left[ {\bf r}(t)  \right]  ({\bf r}'', {\bf r}' ; t'',t')  
\nonumber \\
& = &
S^{(0)}   
\left[ {\bf r}(t)  \right]  ({\bf r}'', {\bf r}' ; t'',t')  
-
\int_{t'}^{t''} 
dt \;
V({\bf r} (t),t)
\;  ,
\label{eq:Hamiltonian_perturb_split}
\end{eqnarray}
where
$
S^{(0)}  \left[ {\bf r}(t)  \right]  ({\bf r}'', {\bf r}' ; t'',t')  
$ 
is the action for the ``unperturbed''
problem whose
solution is already known,
while
$
\tilde{S}  \left[ {\bf r}(t)  \right]  ({\bf r}'', {\bf r}' ; t'',t')  
$ 
represents the perturbation 
in the form of additional interactions 
$
V({\bf r},t )
$.
In what follows, these interactions and the 
unperturbed Hamiltonian will be assumed to be time independent---the generalization
for time-dependent potentials is straightforward and discussed in
Ref.~\cite{cam:pi_bs} and will not be pursued in this paper.
The expansions are generated through 
the Taylor series of  $e^{i\tilde{S}/\hbar}$ in
Eq.~(\ref{eq:propagator_QM}), combined  with
a proper rearrangement of the time lattice that defines the
path integral~\cite{cam:pi_bs,discrete_lattice}.
This procedure leads to the following 
infinite perturbative series for the propagator,
\begin{eqnarray}
K_{D}({\bf r}'', {\bf r}' ; T)
& = &
\sum_{n=0}^{\infty}
\prod_{\alpha=1}^{n} \left[
 \int
d^{D} {\bf r}_{\alpha} \,
\frac{   V({\bf r}_{\alpha})  }{i \hbar} 
\right]
\nonumber 
\\
& \times &
\int_{t'}^{t''} dt_{n} 
\int_{t'}^{t_{n}} dt_{n-1} 
\ldots
\int_{t'}^{t_{2}} dt_{1} 
\,
\prod_{\beta=0}^{n} 
\left[
K^{(0)}_{D}({\bf r}_{\beta + 1}, {\bf r}_{\beta} ; 
t_{\beta+1}-t_{\beta} )  \right]
\;   ,
\label{eq:propagator_perturb_expansion}
\end{eqnarray}
where it is implicitly understood that, at every order $n$,
$t_{0} \equiv t'$, $t_{n+1} \equiv t''=t'+T$, 
${\bf r}_{0} \equiv {\bf r'}$,
and 
${\bf r}_{n+1} \equiv {\bf r''}$;
its corresponding energy Green's function becomes
\begin{equation}
G_{D} ({\bf r''},{\bf r'} ; E) =
\sum_{n=0}^{\infty}
\,
\prod_{\alpha=1}^{n} 
\left[ \int_{0}^{\infty}
d^{D} {\bf r}_{\alpha} 
 V( {\bf r}_{\alpha})  \right]
\,
\prod_{\beta=0}^{n} \left[ 
G_{D}^{(0)} ({\bf r}_{\beta +1},{\bf r}_{\beta};E) 
  \right]
\;  .
\label{eq:GreenF_perturb_expansion}
\end{equation}
In Eqs.~(\ref{eq:propagator_perturb_expansion}) and
(\ref{eq:GreenF_perturb_expansion}),
$K^{(0)}_{D}({\bf r''},{\bf r'} ; T) $
and
$G^{(0)}_{D}({\bf r''},{\bf r'} ; E) $
stand for the unperturbed propagator and energy Green's function, respectively.
In particular,
when the unperturbed problem
is the free-particle case, as we will assume for the remainder of this paper,
the results of Appendix~\ref{sec:propagator_and_GreenF_from_PI}
[specifically Eq.~(\ref{eq:free_particle_GF_not_rescaled})] imply that
\begin{equation}
G^{(0)}_{D}({\bf r''},{\bf r'} ; E_{k}) 
=
  \frac{2M}{\hbar^{2}}
\,
{\mathcal G}^{(+)} _{D} ({\bf R}; k) 
\;  ,
\label{eq:free_particle_GF_rescaled}
\end{equation}
where 
\begin{equation}
k
=
\frac{ \sqrt{2ME_{k}}}{
\hbar}
\;  ,
\; \; \; \; \; \; \; \; \; \; \; \; 
{\bf  R}={\bf  r} - {\bf  r}'
\;  ,
\end{equation}
and
the rescaled
Green's function
${\mathcal G}^{(+)} _{D} ({\bf R}; k) $ is explicitly given by the closed analytical
expression 
\begin{equation}
{\mathcal G}^{(+)} _{{D}}
({\bf R};k)
 =
-
\,
\frac{i}{4}
\left( \frac{k}{2\pi R} \right)^{\nu}
H^{(1)}_{\nu}(kR)
\; ,
\label{eq:free_particle_GF}
\end{equation}
with
\begin{equation}
 \nu= \frac{D}{2} - 1
\;
\label{eq:dimensionality_parameter}
\end{equation}
and
$H^{(1)}_{\nu}(\xi)$
being the Hankel function of the first kind and 
 order $\nu$. 
As shown in Appendix~\ref{sec:GreenF_from_diff_eq},
Eq.~(\ref{eq:DD_GF2_APPENDIX}), 
the function~(\ref{eq:free_particle_GF})
is identical to the $D$-dimensional free-particle
causal energy Green's function (with outgoing boundary conditions),
in agreement with the 
general result~(\ref{eq:GreenF_operator}).
Finally, in the analysis of the bound-state sector of the theory,
the analytic continuation
[cf.~Eqs.~(\ref{eq:DD_GF2_mod_APPENDIX}) and (\ref{eq:DD_GF_vs_DD_GF_mod})]
\begin{equation}
{\mathcal K}_{{D}}
({\bf  R};\kappa)
 =
{\mathcal G}^{(+)} _{D} 
({\bf  R}; k=i \kappa)
=
- \frac{1}{2 \pi}
\left( \frac{\kappa}{2\pi R}
\right)^{\nu}
K_{\nu}(\kappa R)
\; 
\label{eq:DD_GF2_mod}
\end{equation}
will become ubiquitous.

\subsection{Central potentials}

For interactions with central symmetry in $D$ dimensions,
an alternative approach is to 
rewrite the propagator in hyperspherical polar
coordinates~\cite{cam:pi_bs,cam:dtI,erd:53},
in terms of which
Eq.~(\ref{eq:propagator_QM})
admits the expansion~\cite{wat:44}
\begin{equation}
K_{D}({\bf r}^{''}, {\bf r}^{'}; T) 
=
\left( r'' r' \right)^{-(D-1)/2}
\,
\sum_{l= 0}^{\infty}
\sum_{m=1}^{d_{l}}
Y_{l m} ({\bf \Omega''})
Y_{l m}^{ *} ({\bf \Omega'})
K_{l +\nu}(r'',r';T)
\;  ,
\label{eq:propagator_partial_wave_exp}
\end{equation}
where 
$\nu$
is defined in Eq.~(\ref{eq:dimensionality_parameter}), 
$Y_{l m} ({\bf \Omega})$ stands for the hyperspherical harmonics, 
and  $d_{l}= (2l+D-2)(l+D-3)!/l!(D-2)!$~\cite{erd:53}.
This is established 
by introducing the necessary hyperspherical angular coordinates
and
repeatedly using  the addition formula~\cite{wat:44}
$e^{iz\cos \psi}
=
(iz/2)^{-\nu} \Gamma (\nu)
\,
 \sum_{l=0}^{\infty} 
(l+\nu)
I_{l+\nu} (iz ) C_{l}^{(\nu)} (\cos \psi )$,
where $I_{p}(z)$ stands for the modified Bessel function
of the first kind,
while $C_{l}^{(\nu)}(x)$ is a Gegenbauer polynomial~\cite{abr:72}.
Equation~(\ref{eq:propagator_partial_wave_exp}) introduces the radial
propagator
\begin{eqnarray}
K_{l+\nu}(r'',r';T)
& = &
\lim_{N \rightarrow \infty}
\left( \frac{M}{2 \pi i \epsilon \hbar} \right)^{N/2}
\,
\prod_{k=1}^{N-1}
\left[
 \int_{0}^{\infty} d r_{k} 
\right]
\, \mu_{l+\nu}^{(N)} [r^{2}] \,
\;  
\nonumber \\
& \times &
\exp \left\{ \frac{i}{\hbar} 
\sum_{j=0}^{N-1}
\left[
\frac{M}{2 \epsilon} 
( {r}_{j+1}
- {r}_{j}  )^{2} -
\epsilon V( r_{j})
\right]
 \right\}
\nonumber
\\ 
& = &
\int
{\mathcal D}  r(t)  \,
\; 
\mu_{l+ \nu} [r^{2}]
\exp \left(
\frac{i}{\hbar} 
\int_{t'}^{t''}
dt
\left\{
\frac{M}{2} \, \left[   
\dot{r} (t)
\right]^{2}
-
V(r(t))
\right\}
\right)
\;   
\label{eq:propagator_QM_spherical_coords_2D_expansion}
\end{eqnarray}
for each angular momentum channel $l$.
Here
a radial functional weight 
$
\mu_{l+\nu}^{(N)} [r^{2}]
$
has been properly
defined~\cite{interdimensional} so that the radial path integral,
supplemented by the condition $r(t) \geq 0$,
can be given a formal continuum representation
in terms of the usual one-dimensional path-integral pseudomeasure 
${\mathcal D}  r(t)  $;
explicitly,
\begin{equation}
\mu_{l+\nu}^{(N)} [r^{2}]
=
\prod_{j=0}^{N-1} 
\left[ 
\sqrt{2 \pi z_{j}} e^{-z_{j}}
I_{l+\nu} (z_{j})
\right]
\;  ,
\end{equation}
where $
z_{j}
=
 M r_{j} r_{j+1}/i \epsilon \hbar
$.
The perturbation expansion for the radial propagator
follows by directly resolving 
Eq.~(\ref{eq:propagator_perturb_expansion})
into its partial-wave components,
\begin{eqnarray}
K_{l+\nu} (r'',r';T) 
& = &
\sum_{n=0}^{\infty}
\, 
\prod_{\alpha=1}^{n} \left[
 \int_{0}^{\infty} 
d r_{\alpha} \,
\frac{   V(r_{\alpha})  }{i \hbar} 
\right]
\nonumber 
\\
& \times &
\int_{t'}^{t''} dt_{n} 
\int_{t'}^{t_{n}} dt_{n-1} 
\ldots
\int_{t'}^{t_{2}} dt_{1} 
\,
\prod_{\beta=0}^{n} 
\left[
K^{(0)}_{l+\nu} (r_{\beta+1},r_{\beta};
t_{\beta+1}-t_{\beta} )  \right]
\;   .
\label{eq:propagator_perturb_expansion_radial}
\end{eqnarray}
Similarly, by performing a Fourier transform
of Eq.~(\ref{eq:propagator_perturb_expansion_radial}) or by resolving
Eq.~(\ref{eq:GreenF_perturb_expansion}) in partial waves,
the radial energy Green's function is given by
\begin{equation}
G_{l+\nu} (r'',r';E) =
\sum_{n=0}^{\infty}
\,
\prod_{\alpha=1}^{n} 
\left[ \int_{0}^{\infty}
d r_{\alpha} 
 V(r_{\alpha})  \right]
\,
\prod_{\beta=0}^{n} \left[ 
G^{(0)}_{l+\nu} (r_{\beta +1},r_{\beta};E) 
  \right]
\;  .
\label{eq:GreenF_perturb_expansion_radial}
\end{equation}
In particular, if the starting (unperturbed) problem
is the free particle, as reviewed  
in Appendix~\ref{sec:propagator_and_GreenF_from_PI}
[from Eq.~(\ref{eq:free_particle_radial_GreenF_APPENDIX})],
the required radial Green's function
becomes 
\begin{equation}
{G}^{(0)}_{l+\nu} (r'',r';E) 
= -
\frac{2M}{\hbar^{2}}
\,
\sqrt{r' r''}
\,
 I_{l+\nu}( \kappa r_{<} ) K_{l+ \nu} (\kappa r_{>} )
\; ,
\label{eq:GreenF_free_particle}
\end{equation}
where $K_{p} (x )$
is the modified Bessel function of the second
kind and order $p$ (Macdonald function)~\cite{abr:72},
while
$r_{<}$ ($r_{>}$) is the smaller 
(larger) of $r'$ and $r''$.

\section{${\boldmath \delta}$-Function Interactions}
\label{sec:Delta-Function_Interactions}

We are now ready to start applying the general framework to 
$\delta$-function potentials. A
$D$-dimensional $\delta$-function interaction centered
at the origin is represented by the potential
\begin{equation}
V({\bf r}) = \sigma \, \delta^{(D)}  (
{\bf r} )
\; ,
\label{eq:delta_interaction}
\end{equation}
where the coupling strength will be conveniently rewritten as 
\begin{equation}
\sigma = - \frac{\hbar^{2}}{2M} \, \lambda
\; ,
\end{equation}
a replacement that will simplify the ensuing dimensional expressions; in particular
$[\lambda] = 1$ (dimensionless)
for $D=2$.

 Many of the equations 
to be shown in this section have already been derived in a 
number of different contexts.
In particular,
Eq.~(\ref{eq:propagator_perturb_expansion})
can be laboriously evaluated by using a recurrence relation and
its particular expression for $D=1$ agrees with the result known
in the literature~\cite{rob:97}.
On the other hand, in this paper we will focus on the 
expansion of the energy Green's function~(\ref{eq:GreenF_perturb_expansion}),
which can be summed as 
a geometric series.
In effect, rewriting~(\ref{eq:GreenF_perturb_expansion})
term by term,
\begin{equation}
G_{D} ({\bf r''},{\bf r'} ; E) =
\sum_{n=0}^{\infty}
G_{D}^{(n)} 
({\bf r''},{\bf r'} ; E) 
\; ,
\label{eq:GF_delta_perturbative_summation}
\end{equation}
each factor
$\int_{0}^{\infty}
d^{D} {\bf r}_{\alpha} \,
 V({\bf r}_{\alpha})  
$
is merely reduced to $\sigma$ and carries the additional instruction that
the replacement 
${\bf r}_{\alpha} \rightarrow {\bf 0}$ be made.
Then,
\begin{equation}
G_{D}^{(n)}
({\bf r''},{\bf r'} ; E) 
=
\sigma^{n}
\,
\left[
G_{D}^{(0)}
({\bf 0},{\bf 0} ; E) 
\right]^{n-1}
\,
G_{D}^{(0)}
({\bf r}'',{\bf 0} ; E) 
\,
G_{D}^{(0)}
({\bf 0},{\bf r}' ; E) 
\;  ,
\label{eq:reduced_GreenF_delta_perturb_expansion_nth_order}
\end{equation}
for $n \geq 1$, while the term of order zero maintains its value
$G_{D} ^{(0)}({\bf r''},{\bf r'} ; E)  $.
Accordingly,
the exact infinite summation of this series 
for finite $\sigma$ leads to the expression~\cite{gro:93}
\begin{equation}
G_{D} ({\bf r''},{\bf r'} ; E) =
G_{D}^{(0)} 
({\bf r''},{\bf r'} ; E) 
-
\frac{
G_{D}^{(0)}  ({\bf r''},{\bf 0} ; E) 
G_{D}^{(0)}  ({\bf 0 },{\bf r'} ; E) }{
G_{D}^{(0)}  ({\bf 0},{\bf 0} ; E) - 1/\sigma }
\;  .
\label{eq:Green_delta}
\end{equation}
Remarkably,
Eq.~(\ref{eq:Green_delta})
summarizes the complete physics of a $\delta$-function perturbation applied
to any problem whose action  is $S^{(0)}$ and exactly described by 
$G_{D}^{(0)} 
({\bf r''},{\bf r'} ; E) $. 

There is only one apparent restriction in the above derivation:
the geometric series involved in the infinite perturbation 
expansion is guaranteed to converge only
for $|\sigma| < |G_{D}^{(0)}  ({\bf 0},{\bf 0} ; E) |^{-1}$;
 however, this restriction can be 
lifted by noticing that the final 
expression~(\ref{eq:Green_delta}) 
provides the desired analytic continuation in the complex $\lambda$ plane.

In this paper
we will consider the particular case when $S^{(0)}$ represents a free-particle action.
Furthermore,  
Eq.~(\ref{eq:Green_delta}) will now be used to study the bound-state sector of the
theory, while  
in Sec.~\ref{sec:scattering} we will reformulate it in the language of the $T$ matrix
for the scattering sector.
The bound states
are immediately recognized from
the pole(s) of Eq.~(\ref{eq:Green_delta});
explicitly,
with the rescaled Green's function
${\mathcal G}^{(+)} _{D} ({\bf R}; k)$
[Eq.~(\ref{eq:free_particle_GF_rescaled})],
 it follows that 
\begin{equation}
{\mathcal G}^{(+)} _{D} \left(
{\bf 0} ; k = i \kappa
\right) 
= - \frac{1}{\lambda} 
\;  ,
\label{eq:delta_BS}
\end{equation}
a condition that implies the existence of at most one bound state.
This bound-state equation can be analyzed in terms of the behavior
of the $D$-dimensional 
 Green's function~(\ref{eq:free_particle_GF})
near the origin.
In particular,
Eq.~(\ref{eq:delta_BS}) displays a unique pole
when ${\mathcal G}^{(+)} _{{D}}
({\bf  0};k)$ 
is finite, in which case the ground-state wave function---derived from the corresponding
residue---becomes [from Eqs.~(\ref{eq:DD_GF2_mod})
and (\ref{eq:Green_delta})]
\begin{equation}
\Psi_{_{\rm (gs)}}
 ({\bf r})
\propto
\left.
G_{D}^{(0)}  ({\bf 0 },{\bf r} ; E_{k})
\right|_{k=i\kappa}
\propto
{\mathcal G}^{(+)} _{{D}}
({\bf  r};k=i\kappa)
\propto
r^{-\nu}K_{\nu}(\kappa r)
 \; ,
\label{eq:delta_wf}
\end{equation}
where $\kappa$ is the solution to Eq.~(\ref{eq:delta_BS}) and
an appropriate normalization constant should be introduced.
Let us now see the details of this technique for particular dimensionalities.

For the one-dimensional case, described by a coordinate $x \in (-\infty,\infty)$,
the radial variable becomes $R= |x|$, while the Green's functions can be 
rewritten in terms of
 $H^{(1)}_{-1/2}(\xi) = (2/\pi \xi)^{1/2} e^{i\xi}$ and 
$K_{-1/2}(\xi) = (\pi/2 \xi)^{1/2} e^{-\xi}$.
Then,
Eqs.~(\ref{eq:free_particle_GF}) 
and (\ref{eq:delta_BS}) 
provide the condition $\kappa =\lambda/2$, which 
implies a ground-state energy 
$E= - (\hbar \lambda/2)^{2}/2M$
and ground-state wave function
$\Psi_{_{\rm (gs)}} (x)= \sqrt{\kappa} e^{-\kappa |x|} $,
in agreement with the known textbook answers~\cite{rob:97}.

We will now focus on the 
two-dimensional case, which exhibits a number of peculiar
features.
For the two-dimensional $\delta$-function interaction,
Eqs.~(\ref{eq:free_particle_GF}) 
and
(\ref{eq:delta_BS}) 
 lead to a divergent expression
and regularization is called for.
This is the problem we announced
earlier and to which we now turn our attention.

\section{Dimensional Regularization}
\label{sec:DR}

In dimensional regularization~\cite{bol:72,lei:75},
one generalizes the expressions from a given physical dimensionality
$D_{0}$ to $D=D_{0}-\epsilon$, with $\epsilon =0^{+}$.
In quantum mechanics, this procedure is
implemented by analytically extending the potential 
$V({\bf r}) \equiv V^{(     { {D_{0}}}  )}({\bf r})$
from $D_{0}$ to $D$ dimensions,
where it takes a generalized form $V^{({D})}({\bf r})   $.  
The goal of this construct 
is to reformulate the problem in terms of
a regularized potential $V^{({D})}({\bf r})   $ that is no longer singular.  
Even though this generalization is somewhat arbitrary, 
the requirement that it be
regular suggests the following straightforward definition
for $V^{({D})}({\bf r})   $~\cite{cam:dtI},
\begin{equation}
\diagram{
& \mbox{\rm real space} & & \mbox{\rm reciprocal space} \cr
\mbox{$D_{0}$  {\rm dimensions},}
        &  V({\bf r}) = V^{(
           { {D_{0}}}  )}({\bf r})
  &  \diagramrightarrow{
       \mbox{${\mathcal F}_{(  { {D_{0}}} )}$}  }{}
          &  \widetilde{V}( {\bf k}) =
           \widetilde{V}^{({{D_{0}}} )}({\bf k})
    \cr
 &  \diagramdownarrow{
        \mbox{${\mathcal D}_{
          {{D_{0} }} \rightarrow
    {{D}}  }$} }{} & &
 \diagramdownarrow{ \mbox{${\mathcal D}_{
                  {{D_{0}}}
          \rightarrow  {{D}} }$} }{}  \cr
\mbox{$D$  {\rm dimensions}, }  &
               V^{({D})}({\bf r})   &
     \diagramleftarrow{
       \mbox{${\mathcal F}^{-1}_{({D})}$}  }{}
         &
\widetilde{V}^{({D})}({\bf k}) =
              \widetilde{V}({\bf k})
    \cr 
}
\;  ,
\label{eq:diagramdef1}
\end{equation}
where, in this commutative diagram,
${\mathcal F}_{( { {D}_{0}})}$ is
the Fourier transform in $D_{0}$ dimensions,
$  {\mathcal F}^{-1}_{({D}) }$
the inverse Fourier transform in $D$ dimensions,
and
${{\mathcal D}_{
     {{D_{0}}} \rightarrow
  {{D}}  }}$ is a shorthand
for dimensional continuation from $D_{0}$ to $D$ dimensions.
This procedure works because the homogeneous property of
Fourier transforms guarantees that if the potential has
a singular homogeneous behavior of degree $-2$, then 
its counterpart
$V^{({D})}({\bf r})   $ is homogeneous of degree $-2+\epsilon$, and therefore
regular~\cite{cam:dtI}.
For example, if one 
attempted a solution of the inverse square potential, naive generalizations 
(such as keeping a $1/r^{2}$ potential and simply changing the dimensionality) would fail, 
but the prescription~(\ref{eq:diagramdef1}) would be successful~\cite{cam:dtI,cam:dtII}.

At the same time,
in order to preserve the physical dimensions of the original theory,
this procedure entails 
changing the dimensions of the
coupling accordingly~\cite{cam:dtI},
\begin{equation}
\sigma
\rightarrow 
\sigma
\, \mu^{\epsilon}
\;  .
\label{eq:coupling_dimensional_extension}
\end{equation}

For the particular case of the $\delta$-function potential, 
its generalization~(\ref{eq:diagramdef1}) from
$D_{0}$ to $D$ dimensions happens to be the ``obvious'' one,
\begin{equation}
            V^{({D})}({\bf r})   
=
   \sigma \, \mu^{\epsilon} \,
\delta^{({D})} ({\bf r})
\;   .
\end{equation}
Assuming that the two-dimensional case is considered
(by selecting $D_{0}=2$),
the regularized bound-state energy condition~(\ref{eq:delta_BS}) can be directly applied,
provided that $D=2-\epsilon$ and the replacement~(\ref{eq:coupling_dimensional_extension})
is made; then,
\begin{equation}
{\mathcal K}_{{D}} ({\bf 0}; \kappa)
=-
\frac{1}{ \lambda \, \mu^{\epsilon} }
\;  ,
\label{eq:delta_eigen_green}
\end{equation}
where 
${\mathcal K}_{{D}}
({\bf  R};\kappa)
$ is defined in Eq.~(\ref{eq:DD_GF2_mod}).
From the small-argument behavior of 
the modified Bessel functions
 of the second kind~\cite{abr:72,cam:dtI,cam:dtII}:
\begin{equation}
  K_{p} (z)
  \stackrel{(z \rightarrow 0)}{\sim}
 \frac{1}{2}
\,
\left[
\Gamma (p)
\left( \frac{z}{2} \right)^{-p}
 +
\Gamma (-p)
\left( \frac{z}{2} \right)^{p}
\right]
\,
\left[ 1 + O(z^{2}) \right]
\;  
\label{eq:mod_Bessel_small_arg}
\end{equation}
(which amounts to a logarithmic 
singularity for $p=0$),
the following condition is obtained~\cite{cam:dtI},
\begin{equation}
\frac{\lambda \, \mu^{\epsilon}}{4\pi}
\left(  \frac{2M}{\hbar^{2}} \,
\frac{|E|}{4\pi}
\right)^{-\epsilon/2}
\Gamma
\left(\frac{\epsilon}{2} \right) = 1
\; ,
\label{eq:delta_eigen2}
\end{equation}
which displays a simple pole at $\epsilon=0$, making the theory singular for 
the two-dimensional unregularized case.

Renormalization proceeds by introducing the running
coupling~\cite{cam:dtI,cam:dtII},
\begin{equation}
\lambda(\epsilon) = 2 \pi \epsilon
\left\{ 1 + \frac{\epsilon}{2}  \,
\left[
g^{(0)}
- \left(  \ln 4\pi - \gamma \right)
\right]
\right\}
\; ,
\label{eq:delta_renormalized_coupling_DR}
\end{equation}
where $\gamma$ is the Euler-Mascheroni constant; 
from Eq.~(\ref{eq:delta_renormalized_coupling_DR}),  the ground-state 
energy becomes
\begin{equation}
E_{_{\rm (gs)}}
=
- 
\frac{\hbar^{2}\mu^{2}}{2M}
\, 
e^{ g^{(0)}  }
\;  .
\label{eq:delta_gs_energy_DR}
\end{equation}
Notice that, as a result of the arbitrariness in the choice of
$g^{(0)}$, 
we have the freedom to subtract, along with the pole,
the term
$  (\ln 4\pi - \gamma)$;
this is  the analog of the
usual modified minimal subtraction ($\overline{\rm MS}$)
scheme~\cite{pes:95}.

Finally, the residue at the pole straightforwardly provides the 
ground-state wave function, according to
Eq.~(\ref{eq:delta_wf}),
\begin{equation}
\Psi_{_{\rm (gs)}}
 ({\bf r})
=
\frac{\kappa}{ \sqrt{\pi} } \, K_{0} (\kappa r)
 \; ,
\label{eq:delta_wf_normalized_renormalized}
\end{equation}
where
\begin{equation}
\kappa
=
\mu  \, e^{ g^{(0)}/2 }
\;  ,
\end{equation}
a result in agreement with the corresponding one within 
 the Schr\"{o}dinger-equation
approach~\cite{cam:dtI,cam:dtII}.
In short, we have reproduced the familiar results:
(i) 
 the unregularized problem
has a singular spectrum
with a unique energy level at $-\infty$;
(ii) regularization lifts  this level
to a finite value;
(iii) renormalization provides a well-defined prescription
that yields
the {\em unique\/} finite ground state of the
two-dimensional $\delta$-function potential.

Most interestingly,
renormalization of the two-dimensional $\delta$-function potential
leads to the emergence of an arbitrary 
dimensional scale, as seen in the above derivation. 
This remarkable phenomenon,
known as dimensional transmutation~\cite{col:73}, 
implies a violation
of the manifest SO(2,1) symmetry of this scale-invariant potential
and amounts to a simple realization of a quantum anomaly~\cite{jac:91}.
A similar symmetry analysis applies to the inverse square 
potential~\cite{jac:72,alf:76}, magnetic monopole~\cite{jac:80},
and magnetic vortex~\cite{jac:90}, and has recently been generalized
to the dipole potential of molecular physics~\cite{cam:dipole}.

\section{Momentum-Cutoff Regularization}
\label{sec:momentum_cutoff}

Equation~(\ref{eq:delta_BS}) determines the bound-state sector of the 
theory, as discussed in Sec.~\ref{sec:Delta-Function_Interactions}.
An alternative regularization technique can be introduced 
by rewriting 
$G_{{D}}
({\bf  0}, {\bf 0};E)$ 
in the momentum representation ($D=2$), 
a procedure that is equivalent to the use of the integral representations
of Appendix~\ref{sec:GreenF_from_diff_eq}. 
In particular, the replacement of
Eq.~(\ref{eq:DD_GF_mod}) 
 in Eq.~(\ref{eq:delta_BS})  leads to
the integral expression
\begin{equation}
\frac{\lambda }{(2\pi)^{{2}}}
\int
\frac{d^{{2}}q}{q^{2}+ (2 M|E|)/\hbar^{2} } = 1
\;  ,
\label{eq:delta_eigen_Fourier}
\end{equation}
where $E= - | E |$
(with $E<0$ for the possible bound states)
and the variable $q$ is a wave number.

If the integral in Eq.~(\ref{eq:delta_eigen_Fourier})
 is computed naively, an infinite result is obtained.
Nevertheless,
if it is generalized to $D$ dimensions and dimensional regularization
is applied, one can immediately reproduce the results
of Ref.~\cite{cam:dtI} and Sec.~\ref{sec:DR}.
Alternatively,
a completely different regularization can be
implemented through a momentum cutoff $\hbar \Lambda$.
In fact, $\Lambda$ can be introduced from the outset,
directly at the level of the momentum integrals 
derived from the path-integral and Green's function formulations.
With this cutoff procedure,
Eq.~(\ref{eq:delta_eigen_Fourier})
can be straightforwardly integrated, yielding
the result~\cite{tho:79,hua:92}
\begin{equation}
\frac{\lambda}{4 \pi} \ln \left(
\frac{\hbar^{2} \Lambda^{2}}{2M |E|}
 + 1
\right) =1
\;  ;
\label{eq:delta_renormalized_coupling_momentum_cutoff}
\end{equation}
this is equivalent to the statement
\begin{equation}
E_{_{\rm (gs)}}
=
- \frac{ \hbar^{2} \Lambda^{2} }{2M}
\,
\frac{1}{e^{4\pi/\lambda} -1}
\sim
- \frac{ \hbar^{2} \Lambda^{2} }{2M}
\,
e^{-4\pi/\lambda}
\, .
\label{eq:delta_gs_energy_momentum_cutoff}
\end{equation}
Equations~(\ref{eq:delta_renormalized_coupling_momentum_cutoff})
and
(\ref{eq:delta_gs_energy_momentum_cutoff})
lead to the same conclusions as in Sec.~\ref{sec:DR}
[cf. Eq.~(\ref{eq:delta_gs_energy_DR})],
provided that $\lambda=\lambda (\Lambda)$ in such a way that $|E|$
remains finite when $\Lambda \rightarrow \infty$;
this condition requires 
that
\begin{equation}
\lambda (\Lambda) =
\frac{ 4 \pi}{  \ln 
\left(  
\Lambda^{2}/\kappa^{2} + 1   
\right)
}
\sim
-
\frac{2 \pi}{
\ln 
\left(
\kappa/\Lambda
\right)
}
\;  .
\label{eq:delta_renormalized_running_coupling_momentum_cutoff}
\end{equation}
In the language of the renormalization group, this behavior leads to a 
Callan-Symanzik
$\beta$ function~\cite{pes:95}
\begin{equation}
\beta (\lambda)
=
\Lambda \frac{d \lambda} {d \Lambda} 
\sim
 -
\frac{\lambda^{2}}{2 \pi}
\;  ,
\label{eq:beta_function}
\end{equation}
which shows the existence of an ultraviolet fixed point at zero coupling strength.

\section{Real-Space Regularization}
\label{sec:real_space_cutoff}

Real-space regularization
 may be viewed
as the most ``physical'' regularization scheme,  inasmuch as it
explicitly modifies the short-distance physics in order to
provide a well-defined problem.
Of course, there are many possible real-space regularization 
schemes. Here, it proves convenient to introduce a
real-space regulator $a$
such that
\begin{equation}
\delta^{(2)}({\bf r}) \sim
\frac{ \delta (r-a)}{2 \pi a}
\;  ,
\end{equation}
where the limit
$a \rightarrow 0$
is understood.
This amounts to the regularized circular $\delta$-function
potential
\begin{equation}
V({\bf r}) = - 
\frac{ \hbar^{2}\, \lambda}{2M}
\,
\frac{ \delta (r-a)}{2 \pi a}
\equiv
g
\,
\delta (r-a)
\;  ,
\label{eq:circular_delta}
\end{equation}
which can be 
dealt with using the techniques developed in this paper.

Due to the central nature of Eq.~(\ref{eq:circular_delta}),
the formalism of Sec.~\ref{sec:framework} can be directly applied
with the goal of obtaining the radial analog of Eq.~(\ref{eq:Green_delta}),
but now
with the support of the $\delta$ function at $r=a$.
This can be accomplished by a procedure 
similar to the one used in Sec.~\ref{sec:Delta-Function_Interactions}:
(i)
rewriting~(\ref{eq:GreenF_perturb_expansion_radial})
term by term,
\begin{equation}
G_{l+\nu} ( r'', r'; E) 
=
\sum_{n=0}^{\infty}
G_{l+\nu}^{(n)} 
 ( r'', r'; E) 
\; ;
\label{eq:GF_delta_perturbative_summation_radial}
\end{equation}
(ii) performing the integrals
at each order to obtain 
\begin{equation}
G_{l+\nu}^{(n)} 
 ( r'', r'; E) 
=
g^{n}
\,
\left[
G_{l+\nu} ^{(0)}( a, a; E) 
\right]^{n-1}
\,
G_{l+\nu} ^{(0)}
( r'', a; E) 
\,
G_{l+\nu} ^{(0)}
(a, r'; E) 
\;  ,
\label{eq:reduced_GreenF_delta_perturb_expansion_nth_order_radial}
\end{equation}
for $n \geq 1$;
and (iii) summing the series, with
the final result~\cite{cam:pi_bs}
\begin{equation}
G_{l+\nu} ( r'', r'; E) 
=
G_{l+\nu}^{(0)} ( r'', r'; E) 
-
\frac{
G_{l+\nu}^{(0)} ( r'', a; E)
G_{l+\nu}^{(0)} ( a, r'; E)  
}{
G_{l+\nu}^{(0)} ( a, a; E)  
 - 1/g }
\;  .
\label{eq:radial_Green_delta}
\end{equation}
Then,
 the bound-state equation reads
\begin{equation}
\frac{ \hbar^{2} }{2M}
\,
{G}^{(0)} _{l+\nu} 
\left(
a,a; E = - \frac{\hbar^{2} \kappa^{2}}{2M}
\right) 
= - \frac{2 \pi a}{\lambda} 
\; ,
\label{eq:radial_delta_BS}
\end{equation}
which, from Eq.~(\ref{eq:GreenF_free_particle}),
is equivalent to
\begin{equation}
 I_{l} (\kappa a)   \,
K_{l} (\kappa a) 
= \frac{2 \pi}{\lambda}
\;  ,
\label{eq:delta_BS_real_space}
\end{equation}
where $\nu=0$ for $D=2$.

Finally,
Eq.~(\ref{eq:delta_BS_real_space})
can be studied with the small-argument behavior of
the modified Bessel functions
of the first kind~\cite{abr:72,cam:dtI,cam:dtII}
 \begin{equation}
I_{p}(z)
 \stackrel{(z \rightarrow 0)}{\sim} 
\left( \frac{z}{2} \right)^{p} [ \Gamma (p+1) ]^{-1}
\,
\left[ 1 + O(z^{2}) \right]
\;  
\end{equation}
and
 of the second kind, Eq.~(\ref{eq:mod_Bessel_small_arg}). 
The resulting analysis
is summarized by the following conclusions.
If a solution were sought for $l \neq 0$,
the regular boundary condition at the origin would not be satisfied,
as follows from the small-argument expansion~(\ref{eq:mod_Bessel_small_arg}).
Therefore,
the boundary condition at the origin 
is {\em only\/} satisfied for $s$ states,
\begin{equation}
l=0
\;  .
\label{eq:delta_s_states}
\end{equation}
Not surprisingly,
the $\delta$-function potential, being of zero range,
can sustain bound states only in the absence of a centrifugal barrier.

For $l=0$ one then obtains the ground-state 
energy condition
\begin{equation}
E_{_{\rm (gs)}}
=
- \frac{ \hbar^{2} \kappa^{2} }{2M}
=
- 
\frac{ \hbar^{2}  }{2M}
\, 
\frac{4
e^{-2 \gamma}}{
 a^{2} }
\,
e^{-4\pi/\lambda}
\;  ,
\end{equation}
where again $\gamma$ is the Euler-Mascheroni constant
[arising from the expansion~(\ref{eq:mod_Bessel_small_arg}) for $p=0$].
Just as before, renormalization
requires the running of the coupling parameter,
\begin{equation}
\lambda (a) = - \frac{2 \pi}{\ln (\kappa a/2) + \gamma}
\;  ,
\label{eq:delta_renormalized_running_coupling_real_space}
\end{equation}
 an expression that reproduces the same results as 
in Secs.~\ref{sec:DR} 
[Eqs.~(\ref{eq:delta_renormalized_coupling_DR})
and
(\ref{eq:delta_gs_energy_DR})]
and \ref{sec:momentum_cutoff}
[Eqs.~(\ref{eq:delta_gs_energy_momentum_cutoff})
and (\ref{eq:delta_renormalized_running_coupling_momentum_cutoff})].
Specifically, with the identification
$a \sim 1/\Lambda$, the running 
couplings~(\ref{eq:delta_renormalized_running_coupling_momentum_cutoff})
and
(\ref{eq:delta_renormalized_running_coupling_real_space}) 
are in exact correspondence up to finite parts, and the
$\beta$ function associated with~Eq.(\ref{eq:delta_renormalized_running_coupling_real_space}) 
coincides with Eq.~(\ref{eq:beta_function}), as expected on physical grounds.

\section{Scattering}
\label{sec:scattering}

In this section we will consider the scattering sector of theory.
First, in Sec.~\ref{sec:scattering_gral_framework} 
the necessary scattering framework will be developed in 
$D$ dimensions---as it relates to the theory of 
Sec.~\ref{sec:framework}---and applied to the unregularized $\delta$-function interaction.
Second, in Sec.~\ref{sec:scattering_renormalization}
the regularization and renormalization
of the two-dimensional case will be analyzed.

\subsection{Derivation of scattering observables from infinite summation 
of perturbation theory}
\label{sec:scattering_gral_framework}

The scattering sector of theory 
is described most compactly by the Lippmann-Schwinger
equations, which we will present in $D$ dimensions.
These equations can be directly
obtained within our path integral formulation, 
starting from Eq.~(\ref{eq:GreenF_perturb_expansion}),
whose right-hand side stands for the real-space representation of the
 symbolic operator relation
\begin{eqnarray}
G_{D} ( E) 
& = &
\sum_{n=0}^{\infty}
\,
G_{D}^{(0)} (E)
\left[
V G_{D}^{(0)} (E)
\right]^{n}
\nonumber 
\\
& = &
G_{D}^{(0)} (E)
+
G_{D}^{(0)} (E)
V G_{D}(E)
\;  ,
\label{eq:Lippman_Schwinger_GreenF}
\end{eqnarray}
in which it was assumed that 
\begin{equation}
V ({\bf r''},{\bf r'}) = \delta^{(D)} ({\bf r''}-{\bf r'}) \, V({\bf r'})
\, 
\end{equation}
for the local interactions considered in this paper.
The Lippmann-Schwinger equation~(\ref{eq:Lippman_Schwinger_GreenF}) for the Green's function
can be rewritten as a corresponding equation for the $T$ matrix 
\begin{eqnarray}
T_{D} ( E) 
& \equiv &
V
+
V G_{D}(E) V
\nonumber 
\\
& = &
V
+
V G_{D}^{(0)}(E) T_{D} ( E) 
\;  .
\label{eq:Lippman_Schwinger_T_matrix}
\end{eqnarray}
Moreover,  Eq.~(\ref{eq:Lippman_Schwinger_T_matrix})
can be evaluated term by term with the counterpart of
Eq.~(\ref{eq:GF_delta_perturbative_summation}), 
\begin{equation}
T_{D} ({\bf r''},{\bf r'} ; E) =
\sum_{n=1}^{\infty}
T_{D}^{(n)} 
({\bf r''},{\bf r'} ; E) 
\; ,
\label{eq:T_matrix_delta_perturbative_summation}
\end{equation}
which leads to the 
recursion relations
\begin{eqnarray}
T_{D}^{(1)} (E) 
&= &
V
\nonumber 
\\
T_{D}^{(n)} (E) 
& = &
V 
G_{D}^{(0)} (E)
T_{D}^{(n-1)} (E) 
\; ,
\label{eq:Lippmann_Schwinger_nth_order}
\end{eqnarray}
where $n \geq 2$ and the first line represents 
the initial condition ($n=1$).
For a $\delta$-function interaction of the form~(\ref{eq:delta_interaction}),
the recursion relations~(\ref{eq:Lippmann_Schwinger_nth_order})
can be applied sequentially or inductively
to prove that, at order $n$ in perturbation theory,
\begin{equation}
T_{D}^{(n)}({\bf r''},{\bf r'} ; E) =
\sigma^{n}
\,
\left[
G_{D}^{(0)}  ({\bf 0},{\bf 0} ; E) 
\right]^{n-1}
\,
\delta^{(D)} ({\bf r''})
\,
\delta^{(D)} ({\bf r'})
\; ,
\label{eq:T_matrix_delta_order_n}
\end{equation}
which implies the bilocal
form for the $T$ matrix
\begin{equation}
T_{D}  ({\bf r''},{\bf r'} ; E) =
\frac{1}{
1/\sigma - G_{D}^{(0)}  ({\bf 0},{\bf 0} ; E) }
\;
\delta^{(D)} ({\bf r''})
\delta^{(D)} ({\bf r'})
\; .
\label{eq:T_matrix_delta}
\end{equation}

Once the general framework is established,
the computation of the elastic scattering 
amplitude 
$f^{(D)}_{k}(\Omega^{({D})}) $
can be implemented in $D$ dimensions using
the familiar results of scattering theory, by studying the asymptotic behavior of the
corresponding causal Green's function
${\mathcal G}_{D}^{(+)}  ({\bf r}'',{\bf r}' ; E) $
[Eqs.~(\ref{eq:free_particle_GF_rescaled})
and  (\ref{eq:free_particle_GF})].
An alternative and more insightful
albeit lengthier approach---completely based on a path-integral
representation of the $S$ matrix
along the lines of Ref.~\cite{ger:80}---will be reported elsewhere.
In what follows,
${\bf k''} \equiv (k'', \Omega^{({D})}  )$, with $\Omega^{({D})}$ being the set of hyperspherical
coordinates associated with the outgoing wave vector ${\bf k''}$. 
According to the usual formulation~\cite{tay:72}
$f^{(D)}_{k}(\Omega^{({D})}) $ is proportional to the on-shell scattering matrix
elements in the momentum representation, defined as
$
\left\langle 
{\bf k}''
|
T_{D}  (E) 
|
{\bf k}'
\right\rangle
$,
with
$|{\bf k}'|= |{\bf k}'' | \equiv | {\bf k}|$ and
 $E \equiv E_{k}= \hbar^{2}k^{2}/2M$; explicitly,
\begin{equation}
f^{(D)}_{k}(\Omega^{({D})}) 
=
- \frac{1}{4 \pi} 
\,
\frac{2M}{\hbar^{2}}
\,
\left( \frac{k}{2 \pi}
\right)^{({D}-3)/2}
\left.
 \int d^{{D}} r'' 
 \int d^{{D}} r' 
e^{i \left(  {\bf  k}' \cdot {\bf  r}'  -{\bf  k}'' \cdot {\bf  r}'' \right) }  
     T_{D}  ({\bf r''},{\bf r'} ; E_{k}) 
\right|_{ |{\bf k}'|= |{\bf k}''| \equiv |{\bf k}| }
\;  ,
\label{eq:DD_LS_asymptotic}
\end{equation}
where ${\bf k'}$ stands for the incident wave vector. 

Thus,
for the contact interaction~(\ref{eq:delta_interaction}),
substitution of the
$T$ matrix~(\ref{eq:T_matrix_delta}) 
in~(\ref{eq:DD_LS_asymptotic}) yields
\begin{equation}
f^{(D)}_{k}(\Omega^{({D})}) 
 = 
-
 \Gamma_{{D}} (k)
\,
\frac{1}{
1/\lambda +
{\mathcal G}^{(+)}_{{D}} ({\bf 0}; k)}
\;  ,
\label{eq:scattering_amplitude_delta_interaction}
\end{equation}
where
\begin{equation}
 \Gamma_{{D}} (k) =
- \frac{1}{4 \pi} \left( \frac{k}{2 \pi}
\right)^{( {D} -3)/2}
\;  .
\label{eq:Gamma_D}
\end{equation}
Equation~(\ref{eq:scattering_amplitude_delta_interaction})
immediately provides conclusions in agreement with the bound-state
sector of the theory for an attractive potential.
Let us now see how this works for $D=1$ and $D=2$.

In the one-dimensional case
 [with $\nu =-1/2$ and
$\Omega^{(1)} \equiv {\rm sgn} (x)$;
see comments in the paragraph after Eq.~(\ref{eq:delta_wf})],
the Green's function~(\ref{eq:free_particle_GF})
becomes the familiar function
$e^{ik|x|}/2ik$ and
the known transmission $T= |1+if_{k}(+)|^{2}= (2k/\lambda)^{2}/[1+(2k/\lambda)^{2}]$
and reflection  
$R= |if_{k}(-)|^{2}= 1/[1+(2k/\lambda)^{2}]$
coefficients are recovered~\cite{flu:74}. 

In the two-dimensional case,
Eq.~(\ref{eq:scattering_amplitude_delta_interaction}) is divergent 
as a consequence of the
small-argument limit of
Eq.~(\ref{eq:free_particle_GF}),
as we will explicitly verify 
next.

\subsection{Renormalization of the scattering sector}
\label{sec:scattering_renormalization}

In this
subsection
we will
apply the procedure introduced
in~\ref{sec:scattering_gral_framework} 
to the two-dimensional $\delta$-function potential.

A crucial point in this approach is that,
once the need for renormalization is identified,
we have to show the compatibility of the renormalization procedures of both sectors
(bound-state and scattering).
This is most easily accomplished by first renormalizing the bound-state sector
and then using the corresponding running coupling
to eliminate the divergences in the scattering sector.
Physically, this procedure endows the theory with {\em predictive power\/}:
if renormalization is implemented to control the behavior
of the ground state,
then the scattering observables are correspondingly and unambiguously fixed,
without any free parameters.

In general,
the technique of the previous subsection
can be implemented for the two-dimensional $\delta$-function
interaction with each one of the regularization methods discussed in this paper.
However,
for the sake of simplicity, we will just derive the scattering cross section 
using dimensional regularization.

In the dimensional-regularization scheme, the $D$-dimensional expressions derived 
below can be viewed as the regularized counterparts of the physical
$D_{0}=2$ case, supplemented by the renormalization of the coupling parameter.
This coupling, $\lambda \mu^{\epsilon}$, can be substituted for the limit of the
Green's function
${\mathcal K}_{{D}} ({\bf r}; \kappa)$ as $r \rightarrow 0$,
according to the 
bound-state condition~(\ref{eq:delta_eigen_green}).
Then,
\begin{equation}
f_{k}^{(D)}
( \Omega^{({D})}   )
=
 \Gamma_{{D}} (k)
\lim_{r \rightarrow 0}
\left[  {\mathcal K}_{{D}} ({\bf r};
\kappa  ) -
{\mathcal G}^{(+)}_{{D}} ({\bf r}; k) \right]^{-1}
\; ,
\label{delta_scatt_ampl_diff}
\end{equation}
where $\kappa$ parametrizes the ground-state energy
$
|E_{_{\rm (gs)}}| = \hbar^{2} \kappa^{2}/2M$.

The limit in Eq.~(\ref{delta_scatt_ampl_diff}) can be obtained from the asymptotic behavior
\begin{equation}
\lim_{r \rightarrow 0}
  {\mathcal K}_{{D}} ({\bf r};
\kappa)
=
-
\frac{1}{ (2 \pi)^{D}}
\int
\frac{d^{{D}}q}{ q^{2}+  \kappa^{2} }
=
-\frac{1}{ ( 4 \pi)^{D/2} } \,
\kappa^{D - 2}
\,
\Gamma \left( 1 - \frac{D}{2} \right)
\; ,
\label{eq:delta_eigen_Fourier_aux_scatt}
\end{equation}
together with the 
analytic continuation
[cf.~Eq.~(\ref{eq:DD_GF_vs_DD_GF_mod})]
\begin{equation}
  {\mathcal G}^{(+)}_{{D}} ({\bf r}; k)
=
 \left.
{\mathcal K}_{{D}} ({\bf r}; \kappa)
\right|_{ \kappa^{2} \rightarrow - (k^{2}+ i \delta) }
\;  ,
\label{eq:DD_GF2_mod_ANALYTIC_CONT}
 \end{equation}
where $\delta=0^{+}$,
whence
\begin{equation}
\lim_{r \rightarrow 0}
  {\mathcal G}^{(+)}_{{D}} ({\bf r}; k)
=
\frac{1}{ (2 \pi)^{D}}
\,
\int
\frac{d^{{D}}q}{k^{2}- q^{2}+i \delta}
=
-
\frac{1}{ ( 4 \pi)^{D/2} } \,
\left(-k^{2}-i \delta \right)^{D/2 - 1}
\Gamma \left( 1 - \frac{D}{2} \right)
\; .
\label{eq:delta_eigen_Fourier_aux_scatt2}
\end{equation}
Finally,  this implies that
\begin{equation}
\lim_{r \rightarrow 0}
 \left[
{\mathcal G}^{(+)}_{{D}} ({\bf r}; k)
-
  {\mathcal K}_{{D}} ({\bf r}; \kappa)
\right]
=
\frac{1}{ ( 4 \pi)^{D/2} } \,
\Gamma \left( 1 - \frac{D}{2} \right)
\left[
\kappa^{ {D} - 2}
- \left( -k^{2} - i \delta \right)^{ {D}/2 - 1}
\right]
\;  ,
\end{equation}
whose limit
$\epsilon \rightarrow 0^{+}$, with $D=2-\epsilon$, becomes
\begin{equation}
\lim_{\epsilon \rightarrow 0}
\lim_{r \rightarrow 0}
 \left[
{\mathcal G}^{(+)}_{{D}} ({\bf r}; k)
-
  {\mathcal K}_{{D}} ({\bf r}; \kappa)
\right]
=
-
\frac{1}{4 \pi}
\left( \ln \kappa^{2}
- \ln k^{2} + i \pi \right)
\;  ,
\label{eq:green_limit}
\end{equation}
because of the identity
$\ln [- (k^{2}+ i \delta) ]=
\ln k^{2} - i\pi$ for the principal branch of the natural logarithm.
Equation~(\ref{eq:green_limit})
finally provides 
 the scattering amplitude~(\ref{delta_scatt_ampl_diff})
as a function of the 
incident energy $E_{k} =\hbar^{2}k^{2}/2M$,
\begin{equation}
f^{(2)}_{k}
( \Omega^{(2)}   )
=
\sqrt{\frac{2\pi}{ k}} \,
\left[ \ln \left( \frac{E_{k}}{ |E_{_{\rm (gs)}} |  } \right)
-i \pi
\right]^{-1}
\;  ,
\label{eq:delta_scattering_amplitude}
\end{equation}
where,
upon taking the limit $\epsilon \rightarrow 0$,
the observable~(\ref{eq:delta_scattering_amplitude}) 
is evaluated in dimension $D_{0}=2$.

Equation~(\ref{eq:delta_scattering_amplitude}) 
is parametrized with the ground-state energy
$E_{_{\rm (gs)}}$, which is a dimensional variable that has replaced the original
dimensionless coupling---the phenomenon of dimensional transmutation~\cite{col:73}.
In addition,
Eq.~(\ref{eq:delta_scattering_amplitude}) 
verifies the expected 
{\em isotropic\/} scattering
of a contact interaction: the two-dimensional $\delta$-function interaction scatters 
only $s$ waves,
with
a scattering phase shift ($l=0$)
\begin{equation}
\tan
\delta^{(2)}_{0}(k)
=
\frac{\pi}{
\ln \left( E_{k}/|E_{_{\rm (gs)}}| \right) }
\;  ,
\label{eq:2D_delta_phase_shift}
\end{equation}
and with phase shifts $ \delta^{(2)}_{l}(k)=0$ for all $l \neq 0$.
These results can be summarized with the diagonal $S$ matrix in the angular
momentum representation,
\begin{equation}
S^{(2)}_{l,l'}(E)
=
\delta_{l 0} \,
\delta_{l' 0}
\;
\frac{
\ln \left( E /|E_{_{\rm (gs)}}| \right) + i \pi
}
{
\ln \left( E /|E_{_{\rm (gs)}}| \right) - i \pi
}
\;  .
\label{eq:2D_delta_scatt_matrix}
\end{equation}
As a final step,  by direct integration of the
differential scattering cross section
$
d \sigma^{(2)}
(E_{k}, \Omega^{(2)} )/
d \Omega_{2}
=
|f^{(2)}_{k}
( \Omega^{(2)}   )|^{2}
$, 
one obtains the total 
scattering cross section
\begin{equation}
\sigma_{2}(E)
=
\frac{4 \pi^{2}}{k}
\,
\frac{1}{
\left[
\ln \left( E/| E_{_{\rm (gs)}}| \right)
 \right]^{2}
+ \pi^{2}
}
\;  .
\label{eq:delta_total_scatt}
\end{equation}

Remarkably,
all the scattering observables
display a logarithmic dependence with respect to
the incident energy, and
with a characteristic scale set by the ground state.
Finally, two relevant checks can be 
made:

(i)
The unique pole of the scattering matrix~(\ref{eq:2D_delta_scatt_matrix})
provides the ground-state energy.

(ii)
Levinson's theorem,
\begin{equation}
\delta^{(2)}_{l}(k=0) 
-
\delta^{(2)}_{l}(k=\infty) = \pi {\mathcal N}_{l}
\; ,
\end{equation}
provides the correct number of bound states: ${\mathcal N}_{0}=1$ and 
${\mathcal N}_{l}=0$ for $l > 0$.

\section{Conclusions}
\label{sec:conclusions}

In summary, we have completed a thorough analysis
of the path-integral derivation for the
two-dimensional $\delta$-function interaction,
including renormalization {\it \`{a} la\/} field theory
and the compatibility of the bound-state and scattering sectors.
Our results are in agreement with the 
previously known ones from the Schr\"{o}dinger-equation approach.
The formalism provided in this paper also allows for generalizations to
arbitrary dimensionalities and further study of the 
three-dimensional case---which is the nonrelativistic
limit of the scalar $\phi^{4}$ field theory
and is relevant for the
question of triviality~\cite{beg:85}.

Our path-integral treatment of the
contact interaction confirms the 
following conclusions.

(i)  The problem of singular potentials and bound states 
is best dealt with by means of the
energy Green's function $G(E)$.

(ii) Infinite summations and resummations of perturbation theory
give the required nonperturbative behavior.

(iii) Proper analytic continuations may be needed in certain regimes.

(iv) Renormalization can be implemented in the bound-state sector to uniquely predict
the scattering observables.

(v) The effective-field-theory program, which  
leads to singular potentials, requires renormalization
in a quantum-mechanical setting,
such as the one presented in this paper.

Extensions of this generic program to other
singular potentials and field theory will be presented elsewhere.

\acknowledgments{This research was supported in part by
an Advanced Research Grant from the Texas
Higher Education Coordinating Board 
and by the University of San Francisco Faculty Development Fund.
We thank Professor Roman Jackiw,
Professor Luis N. Epele, Professor Huner Fanchiotti,
and Professor Carlos A. Garc\'{\i}a Canal
for stimulating discussions.}

\appendix

\section{Free-Particle Propagator and Green's Function within the Path-Integral Formulation}
\label{sec:propagator_and_GreenF_from_PI}

For a free particle,
the integrals involved 
in Eq.~(\ref{eq:propagator_QM2}) can be directly performed.
The result is 
exact even before taking the large-$N$ limit, as it is independent of the order $N$
of the partition of the time lattice,
\begin{equation}
K^{(0)}_{D}({\bf r}^{''}, {\bf r}^{'}; T) 
=
\left( \frac{M}{2 \pi i T \hbar} \right)^{D/2}
\,
\exp \left[
\frac{i
M
( {\bf r}''
- {\bf r}')^{2} }{2 \hbar T}
\right]
\;  .
\label{eq:free_particle_propagator_QM2}
\end{equation}

The corresponding energy Green's function can be computed from 
Eq.~(\ref{eq:GreenF_from_Fourier_transform}), so that
\begin{equation}
G_{D}^{(0)} ({\bf r''},{\bf r'};E) 
=
\frac{1}{i\hbar}
\,
\left( \frac{M}{2 \pi i \hbar} \right)^{D/2}
\,
{\cal I}_{\nu} 
\left(
\tau(R,E),\omega(E)
\right)
\; ,
\label{eq:GreenF_from_Fourier_transform_explicit_calculation}
\end{equation}
where~\cite{gra:00} 
\begin{eqnarray}
{\cal I}_{\nu} (\tau,\omega)
& =  &
\int_{0}^{\infty} 
dT \,
T^{-\nu-1}
\exp 
\left\{
i 
\omega 
\left[
T + \frac{\tau^{2}}{T}
\right]
\right\}
\nonumber
\\
& = &
i \pi \tau^{-\nu}
\exp
\left( 
i \frac{\nu \pi}{2}
\right)
\,
H_{\nu}^{(1)} 
\left(
2 \omega \tau
\right)
\; ,
\label{eq:AUX_GreenF_from_Fourier_transform_explicit_calculation}
\end{eqnarray}
in which
$\nu$ is given by Eq.~(\ref{eq:dimensionality_parameter}),
$\omega (E)= E/\hbar$, and
$\tau(R,E) $ is defined from $\omega (E) [ \tau(R,E) ]^{2} = MR^{2}/2\hbar$,
 with  ${\bf  R}={\bf  r} - {\bf  r}'$.
As  a result, rewriting
$2 \omega \tau = kR$, the Green's function becomes
\begin{equation}
G_{D}^{(0)} ({\bf r''},{\bf r'};E) 
 =
-
\,
\frac{i}{4}
\,
  \left(
\frac{2M}{\hbar^{2}}
\right)
\,
\left( \frac{k}{2\pi R} \right)^{\nu}
H^{(1)}_{\nu}(kR)
\; ,
\label{eq:free_particle_GF_not_rescaled}
\end{equation}
which is equivalent to Eqs.~(\ref{eq:free_particle_GF_rescaled})
and  (\ref{eq:free_particle_GF}).
Finally, the analytic continuations~(\ref{eq:DD_GF2_mod})
and
(\ref{eq:DD_GF2_mod_ANALYTIC_CONT})
of the Green's function
to negative energies---needed for the bound-state sector---can be straightforwardly obtained
with the relation~\cite{abr:72} 
\begin{equation}
K_{\nu}(\mp iz) =\pm \frac{\pi i}{2} e^{\pm i \pi \nu/2}
H^{(1,2)}_{\nu}(z)
\; .
\label{eq:Hankel_MacDonald_connection}
\end{equation}

The corresponding expressions for the free particle in hyperspherical coordinates
can be established from Eq.~(\ref{eq:propagator_QM_spherical_coords_2D_expansion}),
which gives
\begin{eqnarray}
K_{l+\nu}(r'',r';T)
& =  &
\lim_{N \rightarrow \infty}
\sqrt{r' r''}
\,
\alpha^{N}
\,
\prod_{k=1}^{N-1}
\left[
 \int_{0}^{\infty} d r_{k}  \, r_{k} 
\right]
\,
\nonumber \\
& \times &
e^{-\alpha(r_{1}^{2} + \cdots r_{N-1}^{2} )}
\,
I_{l+\nu}( \alpha r_{0} r_{1})
\ldots
I_{l+\nu}( \alpha r_{N-1} r_{N})
\;  ,
\label{eq:free_particle_radial_propagator_APPENDIX}
\end{eqnarray}
where $\alpha = M/i\hbar \epsilon$. 
Equation~(\ref{eq:free_particle_radial_propagator_APPENDIX})
can be evaluated recursively by repeated application of Weber's second exponential 
formula~\cite{wat:44}, with the result~\cite{gro:98}
\begin{equation}
K_{l+\nu}(r'',r';T)
=
\frac{M}{i\hbar T}
\,
\sqrt{r' r''}
\,
\exp 
\left[
\frac{iM}{2 \hbar T} 
( r'^{2} +  r''^{2})
\right] 
\,
I_{l+\nu}
\left(
\frac{Mr'r''}{i\hbar T}
\right)
\;  ,
\end{equation}
whence the Green's function 
becomes (with the symbol $a = M/2\hbar$)
\begin{equation}
G_{l+\nu}(r'',r';T)
=
-
\frac{2a}{\hbar}
\sqrt{r' r''}
\int_{0}^{\infty}
\frac{dT}{T}
\,
I_{l+\nu} \left( \frac{2 a \, r'' r'}{iT} \right)
\,
\exp 
\left\{ 
i
\left[
\frac{(E+ i 0^{+} )}{\hbar} T
+
a \left( r'^{2}+r''^{2} \right)
\frac{1}{T}
\right]
\right\}\;  ,
\label{eq:free_particle_radial_GreenF_APPENDIX}
\end{equation}
which can integrated in closed form
with the substitution $u=1/T$~\cite{gra:00b}
to yield Eq.~(\ref{eq:GreenF_free_particle}).

\section{Free-Particle Green's Functions from Operator Formulation}
\label{sec:GreenF_from_diff_eq}

The connection between the path-integral and operator formulations of the
Green's function can be established by means of Eq.~(\ref{eq:GreenF_operator}).
In particular,
the latter can be recast into the form of
a Green-Helmholtz equation,
which we will consider next.
For the free-particle case, which
applies to
 unbounded space, translational invariance implies that
$G_{{D}} ({\bf  r}, {\bf  r}'; E_{k})$
is only a function of 
${\bf  R}={\bf  r} - {\bf  r}'$, and the rescaling~(\ref{eq:free_particle_GF_rescaled})
can be defined.
Then,
the Green-Helmholtz equation reads
\begin{equation}
\left[
 \nabla_{{\bf  R},{D}}^{2} +
k^{2}
  \right]
{\mathcal G}_{{D}} ({\bf  R};k)
=
\delta^{({D})}
({\bf  R} )
\; .
\label{eq:Helmholtz}
\end{equation}
As is well known,
its Fourier transform
$\widetilde{\mathcal G}_{{D}}({\bf  q};k)
= (k^{2}-q^{2})^{-1}$
leads to an ill-defined  integral expression that needs to be
evaluated by an additional prescription defining the boundary conditions at 
infinity;
for outgoing $(+)$ and incoming $(-)$ boundary conditions,
\begin{eqnarray}
{\mathcal G}^{(\pm)}_{{D}}({\bf  R};k)
& = &
\int \frac{ d^{{D}} q }{ (2 \pi )^{{D}} }
\frac{ e^{ i {\bf  q} \cdot {\bf  R} }}{k^{2}-q^{2}\pm i\delta}
\nonumber \\
& = &
 (2\pi)^{-{D}/2} R^{-({D}/2 - 1)}
\int^{\infty}_{0}
\frac{q^{{D}/2}
J_{{D}/2 - 1}
(q R)}{k^{2}-q^{2} \pm i \delta}dq
\;  ,
\label{eq:DD_GF}
\end{eqnarray}
where $\delta=0^{+}$.
The energy dependence of Eq.~(\ref{eq:GreenF_operator})
shows that ${\mathcal G}^{(+)}_{{D}}({\bf  R};k)$
is the function that reproduces the correct behavior of the path-integral 
expression~(\ref{eq:GreenF_from_PI}).

Similarly,
analytic continuation to negative energies 
leads to the
modified
 Green-Helmholtz equation,
\begin{equation}
\left[
 \nabla_{{\bf  R},{D}}^{2} -
\kappa^{2}
  \right]
{\mathcal K}_{{D}}({\bf  R};\kappa)
=
\delta^{({D})}
({\bf  R} )
\; ,
\label{eq:Helmholtz_mod}
\end{equation}
where the analog of the rescaling~(\ref{eq:free_particle_GF_rescaled})
should be considered.
The 
Fourier transform 
of ${\mathcal K}_{{D}}({\bf  R};\kappa)$
can be derived from Eq.~(\ref{eq:Helmholtz_mod}),
i.e.,
$\widetilde{\mathcal K}_{{D}}({\bf  q};\kappa)
= -(q^{2}+\kappa^{2})^{-1}$, a result that can be finally inverted to give
\begin{eqnarray}
{\mathcal K}_{{D}}({\bf  R};\kappa)
& = & -
\int \frac{ d^{{D}} q }{
(2 \pi )^{{D}} }
\frac{ e^{ i {\bf  q} \cdot {\bf  R} }}{q^{2}+\kappa^{2}
}
 \nonumber  \\
& = & -
 (2\pi)^{-{D}/2}
R^{-({D}/2 - 1)}
\int^{\infty}_{0}
\frac{q^{{D}/2}
J_{{D}/2 - 1}
(q R)}{q^{2}+\kappa^{2} } \, dq
\;  .
\label{eq:DD_GF_mod}
\end{eqnarray}
Equation~(\ref{eq:DD_GF_mod})  can be evaluated
in terms of the modified Bessel  
function of the second kind $K_{\nu} ( \kappa R)$, of order 
$\nu=D/2 - 1$,
i.e.~\cite{gra:00c},
\begin{equation}
{\mathcal K}_{{D}}
({\bf  R};\kappa)
 =
- \frac{1}{2 \pi}
\left( \frac{\kappa}{2\pi R}
\right)^{\nu}
K_{\nu}(\kappa R)
\;  .
\label{eq:DD_GF2_mod_APPENDIX}
\end{equation}

Furthermore,
Eq.~(\ref{eq:DD_GF})
can be expressed in terms of Hankel functions
of order $\nu$;
for example,
Eq.~(\ref{eq:Helmholtz}) can be obtained from (\ref{eq:Helmholtz_mod})
with the replacement $\kappa= \mp i k$,
which provides the solution
\begin{equation}
{\mathcal G}^{(\pm)}_{{D}}({\bf  R};k)
=
{\mathcal K}_{{D}}({\bf  R};\kappa= \mp ik)
\;  ,
\label{eq:DD_GF_vs_DD_GF_mod}
\end{equation}
and the choice of signs amounts to the choice of boundary conditions
at infinity or the $i\delta $ prescription.
From the identity~(\ref{eq:Hankel_MacDonald_connection}), it follows
that
Eq.~(\ref{eq:DD_GF_vs_DD_GF_mod}) acquires the form
\begin{equation}
{\mathcal G}^{(\pm)}_{{D}}
({\bf  R};k)
 =
\mp \frac{i}{4}
\left( \frac{k}{2\pi R} \right)^{\nu}
H^{(1,2)}_{\nu}(kR)
\; ,
\label{eq:DD_GF2_APPENDIX}
\end{equation}
a result that
reduces to familiar expressions for $D=1,2,3$.


\begin{thebibliography}{99}
\bibitem{discrete_lattice}
 In the time-lattice version of  
expression~(\ref{eq:propagator_perturb_expansion}),
at order $n$ in perturbation theory, one introduces
$n+1$ subintervals $T_{\alpha}$---with each subinterval
partitioned in the usual way into $N_{\alpha}$ parts.
Notice the extra integrations with respect to time in 
Eq.~(\ref{eq:propagator_perturb_expansion}) [compared with
Eq.~(\ref{eq:GreenF_perturb_expansion})],
which arise when separating the end points of each subinterval.
\end{thebibliography}
\end{document}